\documentclass[epj]{svjour}

\usepackage{graphics}
\usepackage{amsmath}
\usepackage{multirow}
\usepackage[utf8]{inputenc}

\begin{document}

\title{
Where do we stand with
a 3-D picture of the proton?
}

\author{Alessandro Bacchetta\inst{1}}

\institute{Dipartimento di Fisica, Universit\`a degli Studi di Pavia, and
  INFN, Sez. di Pavia \\
 via Bassi 6, 27100 Pavia, Italy
}

\date{Received: date / Revised version: date}

\abstract{
We are on the way to obtaining multi-dimensional ``pictures'' of the
proton. The field is bursting with activities, both from the theoretical and
experimental side.  
A brief selection of important achievements of the last years and open challenges for the future is
presented. They are already well documented in the various reviews of this
Special Issue, but they are gathered here in a more condensed way for
convenience, together with some additional remarks. 
The choice of the items included in this short overview is far from being
complete and represents the view of the author. 
\PACS{
  {12.38.-t}{Quantum chromodynamics} \and
  {13.60.-r}{Photon and charged-lepton interactions with hadrons} \and
  {13.88.+e}{Polarisation in interactions and scattering}
     } 
} 

\maketitle

\section{Why 3-D maps are interesting?}

This Special Issue is entirely devoted to the description of the state of the
art of the field of 3-D proton ``maps.'' It gives an impressive overview,
both wide and deep, which perfectly conveys the idea
of how vibrant this field of research is. Reading the excellent reviews of
this Special Issue, it is unavoidable to be surprised by the balanced mixture of
theoretical and experimental advances, by the long list of achievements and
the even longer list of to-dos. The field can be rightfully considered as
{\em expanding}.  

What are the characteristics that make this field of research flourish in
these years?  

First of all, this field is focused on one of the ``core missions'' of Physics:
understanding the fundamental
constituents 
of matter.
 
This question has changed its meaning in the course
of the last century. We changed our perspective on the meaning of
``fundamental,'' ``constituent,'' and even ``matter.''

For what concerns the word ``matter,'' we discovered the existence of
dark matter and possibly even of dark energy. Normal matter around us is just
a small fraction of the total matter of the universe. Our curiosity urges us
to understand dark matter and dark energy, but this does not diminish our
interest in normal matter, which remains of
particular relevance for our life and existence. 


Concerning the word ``constituents,'' one century ago we thought that the
constituents of matter were electrons, nuclei, photons. Nowadays the list of
elementary particles includes three families of leptons, three families
of quarks, photons, gluons, weak bosons, the Higgs boson, and all the
antiparticles. A good
number of these particles is never actually ``seen'' in nature. We see their
indirect effects. 

For what concerns the word ``fundamental,'' nowadays at the fundamental level
of our description of Nature 
we place quantum
fields, which have no underlying components and are elementary.
To the best of our knowledge,
electrons are elementary. 
%
Hadrons are not elementary quantum fields, 
they do not appear in the Lagrangian of the Standard Model,
they emerge as bound states of quarks and gluons. 
Hadrons are in fact the
smallest non-elementary structures of the universe: understanding them is the
most basic example of going from an underlying layer of ``simple''
constituents to an emerging layer of 
``complex'' structures.

Hadrons are not elementary, but they are fundamental. 
We can consider atoms to be the fundamental building
blocks in Chemistry, cells are the fundamental building blocks in
Biology. In the same way, protons 
and neutrons can be rightfully counted among the fundamental constituents
of matter.

In short, when studying the structure of protons,
neutrons, and hadrons in general, we are focusing our attention on 
{\em a few} {\em non-elementary} constituents of {\em a small fraction} of
matter. However, this statement can be turned into a different one: 
we are focusing our
attention on the most relevant fundamental constituents of our world's matter.


A second crucial characteristic of this field of research 
is that it is an {\em interface} field. Interface
between theory and experiment (and computation), between high-energy and
nuclear physics, perturbative and nonperturbative, elementary
and complex. 

For instance, the study of the 1-D structure of the proton, encoded in
collinear PDFs,  started in the Seventies at
facilities sitting at the edge of the energy frontier. The emphasis at that
time was understanding the structure of the proton and QCD. 
In the following decades, predictions based on QCD factorization theorems -- in
particular scaling violations and DGLAP evolution equations -- became one of
the most important and successful tests of perturbative QCD. 
In the last decades, PDFs are
studied not only because of the information they provide about QCD, but mainly
because they are useful to access and understand other aspects of the physics
of the Standard Model and beyond. This is the natural destiny of many
scientific discoveries: first we {\em understand}, then we {\em use}. 
First we understood
how to define and extract PDFs, then we used (and still use) them to make
predictions and look for new physics. The two steps are however not always
clearly separated, nor strictly consequential. For instance, very often we use
information without completely understanding it. And using it, we often
improve our understanding. To come again to the example of PDFs, we can claim
that we understand how to extract PDFs in a reliable way (at least for certain
parton flavors and in certain range of $x$). But we have to admit that we do not
understand how the details of PDFs arise from QCD, since we are not able to
solve QCD in the nonperturbative regime. We are using something that we do not
fully understand.

What is the meaning of {\em understanding} the structure of the proton?
A possible answer could be: 
being able to encode our knowledge of the proton in some formulas or software 
that is able to describe what we have measured about the proton 
and make predictions for
something that we have not measured yet.
This task is extremely difficult, but not hopeless. We are already able to
obtain some excellent results from lattice QCD. They are however limited to
certain 
aspect of proton physics and only indirectly related to 3-D
pictures. Furthermore, in order to prove itself to be really useful,
lattice QCD should be able to make predictions that can be tested against new
measurements. The field of 3-D pictures may become the best arena to
develop and test lattice QCD.

Maybe understanding the proton's structure will turn out to be too an ambitious
goal. Nevertheless, 
the amount of information we are gathering with 3-D proton studies is so
large that we will certainly find ways to use it. We will partially follow the
footprints of 1-D studies and use our knowledge to make predictions for
processes involving hadrons. 
For most observables, the
knowledge of the 1-D structure of the hadrons is enough. 
But for some observables, the
details of the 3-D structure are necessary. 
 
Stretching our imagination even further, we could wonder if there will ever be
some application based on the detailed knowledge of the structure of the
proton. This is obviously a very speculative topic. The energies ``stored'' in
nucleons are hundreds of times larger than those stored in nuclei. 
To tap this source of energy, we would need to be able to
transform one hadron to another. It is not something that we can do with
natural elements, therefore it is unlikely that we can produce energy using
QCD. However, in principle it could be possible to use QCD to store
energy. More generally, the closest we get to ``using'' QCD are applications
such as hadron therapy and nuclear fusion. Neither of them make use of
the knowledge of the internal structure of the proton. It is therefore at
present very difficult to foresee any application of this knowledge. The same,
however, can be said of the Higgs boson and of gravitational waves, and yet
they remain strong driving forces for progress in science and elsewhere.

In the next sections, a brief selection of important achievements of the last years
and open challenges for the future is
presented. They are already well documented in the various reviews of this
Special Issue, but it is convenient to partially repeat them also
here. However, the
list of achievements and open challenges is so long that it is impossible to be
complete. The choice represents the view of the author. Also, the list of
references is necessarily incomplete and the reader is invited to look into the
individual reviews for a more extended bibliography.

\section{TMDs}

From the point of view of TMDs, we have accomplished gigantic steps
forward in the last ten years. 

TMD factorization, universality, and evolution
properties have been placed on a firm theoretical ground, thanks to important
contributions from several groups (see, e.g.,
Refs.~\cite{Ji:2004wu,Collins:2011zzd,Aybat:2011zv,GarciaEchevarria:2011rb,Echevarria:2012js}
and the reviews by Rogers and Diehl in this Special Issue~\cite{Rogers:2015sqa,Diehl:2015uka}). 
We have established
that T-odd distribution functions such as the Sivers function can exist~\cite{Brodsky:2002cx}. 
We discovered that TMDs, in particular T-odd ones, get modified when going
from a process to another.
The origin of these differences rests on the structure of the Wilson
lines included in the proper definition of TMDs~\cite{Ji:2002aa,Belitsky:2002sm}. In practice, this results in
a sign difference between semi-inclusive DIS and Drell-Yan~\cite{Collins:2002kn}.
More complicated color
prefactors were supposed to appear in other processes, for instance in
proton-proton collisions to
hadrons~\cite{Bomhof:2004aw,Bacchetta:2005rm}. 
However, it was later demonstrated that in
these processes it
is not possible to disentangle the Wilson line structure and factorization is
thus broken~\cite{Collins:2007nk,Rogers:2010dm,Rogers:2013zha}. Fragmentation functions are not affected by universality
problems, therefore no differences are expected going, for instance, from
semi-inclusive DIS to $e^+ e^-$ annihilation~\cite{Collins:2004nx}. TMD evolution equations were
derived in a clear way, and the connection with the literature of
transverse-momentum resummation was made manifest. At least for the
unpolarized TMD case, the elements involved in the evolution equations are now
known up to Next-to-Next Leading Logarithmic accuracy (see, e.g, Ref.~\cite{Echevarria:2012pw}). 

From the experimental point of view, a vast amount of data coming from
semi-inclusive DIS has been collected, thanks to measurements performed mainly
at HERMES and COMPASS (see also the reviews by Avakian, Bressan, Contalbrigo
and Prokudin, Boglione~\cite{Boglione:2015zyc} in this Special Issue). HERMES played a truly pioneering role in these
advances, with the first measurements of single transverse-spin asymmetries in
DIS~\cite{Airapetian:2004tw}, closely followed by COMPASS~\cite{Alexakhin:2005iw}, which keeps delivering new data also today.
The driving interest initially was to measure transverse-spin dependent
azimuthal modulations,
with the goal of accessing the transversity distribution
function. Longitudinally polarized azimuthal asymmetries have been measured
long before, as well as unpolarized observables. Gradually, more observables
were taken into consideration by HERMES,
COMPASS, and JLab experiments (see, e.g., Refs.~\cite{Mkrtchyan:2007sr,Osipenko:2008rv,Airapetian:2009ae,Airapetian:2009ti,Airapetian:2010ds,Avakian:2010ae,Alekseev:2010rw,Huang:2011bc,Qian:2011py,Airapetian:2012ki,Airapetian:2012yg,Adolph:2013stb,Airapetian:2013bim,,Allada:2013nsw,Zhang:2013dow,Adolph:2014pwc,Parsamyan:2015dfa}). We have measurements, at least in preliminary
form, of all 16 azimuthal modulations in semi-inclusive DIS, for proton and
deuteron targets, and for pion, kaons, and protons in the final
state. Together with this, experiments also measured
dihadron asymmetries and inclusive pion production asymmetries.

In the meanwhile, a complementary effort to measure observables in $e^+e^-$
collisions started and led to crucial measurements both by the BELLE and BABAR
collaborations, followed also by the BESIII collaboration~\cite{Seidl:2008xc,Vossen:2011fk,TheBABAR:2013yha,Ablikim:2015pta}. 

From the side of $pp$ collisions, some measurements related to TMD studies
appeared in Drell-Yan (Fermilab) and in specific processes with jets or hadrons in the
final state (RHIC). Most of the measurements in $pp$ collisions dealt with
single-spin asymmetries with only one hadron or jet in the final state, which
are indirectly related to TMDs. 

\renewcommand{\arraystretch}{1.5}
\begin{table}
\begin{center}
\begin{tabular}{c|c|c|c|c|c|}
\multicolumn{2}{c}{}&\multicolumn{4}{c}{quark pol.}\\
\cline{3-6}
\multicolumn{2}{c|}{}& U & L & \multicolumn{2}{c|}{T} \\ 
\cline{2-6}
\multirow{3}{*}{\rotatebox{90}{nucleon pol.}} &{U} &
{ $\boldsymbol{f_1}$}   & & \multicolumn{2}{c|}{ $h_{1}^\perp$} \\
\cline{2-6}
& {L} & &{ $\boldsymbol{g_{1}}$} & \multicolumn{2}{c|}{ $h_{1L}^\perp$} \\
\cline{2-6}
& {T} & { $f_{1T}^\perp$} &  { $g_{1T}$} &
{ $\boldsymbol{h_{1}}$} & {$h_{1T}^{\perp}$} \\
\cline{2-6} 
\end{tabular}
\end{center}
\caption{Twist-2 Transverse-Momentum-dependent Distribution functions (TMDs).
The U,L,T correspond to unpolarized, longitudinally polarized and transversely
polarized nucleons (rows) and quarks (columns). Functions in boldface survive
transverse momentum integration.}
\label{t:TMDtable}
\end{table}

Table \ref{t:TMDtable} shows all the eight twist-2 TMDs, with their dependence on
polarization. 
From the phenomenological point of view, parameterizations have been presented
for unpolarized TMD PDFs ($f_1$ in the table) and fragmentation function,
Sivers function ($f_{1T}^{\perp}$),
Boer-Mulders function ($h_1^{\perp}$), transversity ($h_1$), ``pretzelosity''
($h_{1T}^{\perp}$), and Collins fragmentation function. Most of
them have been worked out in the  so-called ``Phase 1''
approach, i.e., an approach with no pQCD corrections (order $\alpha_S^0$) (see,
e.g., Refs.~\cite{Collins:2005ie,Vogelsang:2005cs,Anselmino:2007fs,Anselmino:2008jk,Anselmino:2010bs,Schweitzer:2010tt,Bacchetta:2011gx,Signori:2013mda,Anselmino:2013lza,Anselmino:2013vqa,Barone:2015ksa}). This
phase, very important for first explorations, is now almost over and it is
being replaced by the ``Phase 2'' approach, where TMD evolution equations are
taken into consideration. 
However, the implementation of these equations
contains some nonperturbative ingredients and some matching procedures that
are not strictly defined by the theory. Different prescriptions are being
used: they should all be equivalent from the theoretical point of view, but it
may be that some are more convenient than others in practical applications
(see, e.g., the discussion in Ref.~\cite{Collins:2014jpa}). Phase-2 fits have started only for the TMDs in the first
column of Tab.~\ref{t:TMDtable} 
(unpolarized TMD $f_1$ and Sivers function $f_{1T}^{\perp}$~\cite{Aybat:2011ta,Anselmino:2012aa,Echevarria:2014xaa}) and for
transversity, $h_1$~\cite{Kang:2015msa}. 

The driving interest to study transverse spin observables was
the goal of studying the transversity distribution through the so-called
Collins effect. This milestone has been achieved in a series of papers that
carried out the analysis in a constantly improving way (see, e.g.,
Refs.~\cite{Anselmino:2007fs,Anselmino:2013vqa,Anselmino:2015sxa}). 
The state-of-the-art
extractions make full use of TMD evolution equations~\cite{Kang:2015msa}. All
extractions so far are in acceptable
agreement with each other and they are also in fair agreement with independent
extractions
based on collinear factorization in dihadron
production (see Refs.~\cite{Bacchetta:2011ip,Bacchetta:2012ty,Radici:2015mwa}
and the review by Pisano, Radici in this Special Issue~\cite{Pisano:2015wnq}). 
In conclusion, 
we can claim
that we are able to extract the transversity distribution, that the up
transversity is sizable in the medium-large $x$ region and positive (the
absolute sign actually cannot be fixed by experiments), that the down
transversity can be large and negative, although its determination is affected
by larger errors compared to the up quark. Nucleon tensor charges (i.e.,
combinations of integrals of transversity distributions) have been estimated,
although they are affected by extrapolation uncertainties:
the extraction in Ref.~\cite{Radici:2015mwa} is in agreement with most recent lattice QCD
estimates for the isovector combination $u-d$~\cite{Green:2012ej,Bali:2014nma,Bhattacharya:2015wna,Abdel-Rehim:2015owa};
the extraction in Ref.~\cite{Kang:2015msa} is estimated at a higher scale
compared to the lattice calculations. Considering the flavors separately, the
extractions of the $u$ tensor charge are typically lower than lattice results. None of the
extractions has taken sea quarks into considerations.

The Sivers function is large and positive for up quarks. For down quarks it
has an opposite sign, but is, somewhat unexpectedly, of the same size as the up
quark~\cite{Anselmino:2012aa,Echevarria:2014xaa}. Up and down sea quark contributions are small in the $x$ region
explored so far, strange quarks are poorly constrained~\cite{Bacchetta:2011gx}. 

Unpolarized TMDs are not yet constrained in a satisfactory way. They are
present in all measurements, therefore we would expect them to be better known
than anything else. But since they are so ubiquitous, it is not sufficient to
describe their qualitative features: some precision is required. In order to make
some simple statements, let us consider the position of the peak in the
distribution $|{k}_T| f(x,{k}_T^2)$ as a measure of the
``width'' of a TMD $f(x,{k}_T^2)$. 
For a TMD with a Gaussian shape, this width is equal to $\sqrt{\langle {k}_T^2 \rangle/2}$. This
particular definition is 
useful because it can be applied to any function, even if not integrable,
which would not be the case for, e.g., the average transverse momentum squared
$\langle {k}_T^2 \rangle$. Present extractions (see, e.g.,
Refs.~\cite{Signori:2013mda,Anselmino:2013lza,Echevarria:2014xaa,D'Alesio:2014vja})
indicate that the width of the
TMDs at low scales, 1-2 GeV, is around 300-500 MeV. The width increases
to more than 1 GeV at the $Z$ mass, due to TMD evolution. 
Data indicate also that
the width is probably increasing as $x$ decreases and there is room for a
strong flavor dependence, even though also a flavor-independent scenario is
not ruled out~\cite{Signori:2013mda}.

Several model calculations of TMDs, GPDs and 
Wigner distributions have been presented in the last years (see references in
the contribution by Burkardt and Pasquini to this Special Issue~\cite{Burkardt:2015qoa})
). They are in general 
able to capture the qualitative
features of form factors and collinear PDFs,
but they are still far from giving a description that is satisfactory from the
quantitative point of view. However, models have been used especially to
illustrate the physics content of TMDs, to predict their qualitative behavior
and to give an estimate of the possible size of unknown observables. The most
important example of the relevance of model calculations has been the proof
that the 
Sivers function can be nonzero~\cite{Brodsky:2002cx}. Models also usually display some nontrivial
relations among TMDs and GPDs~(see, e.g., Refs.~\cite{Efremov:2002qh,Burkardt:2003yg,Meissner:2007rx,Pasquini:2008ax,Bacchetta:2008af,Bourrely:2010ng,Avakian:2010br,Efremov:2010mt,Lorce:2011zta,Muller:2014tqa}). Some of these relations are broken in
perturbative QCD. However, 
they could still be valid approximations at the threshold
between the nonperturbative and perturbative regimes. As such, they could be
useful conditions to guide TMD parameterizations.

In spite of these impressive results, the formalism has to be 
tested to a higher level of
precision than currently done. 
It is not easy to find a clear-cut way to check the validity of the
formalism. Even in the case of collinear PDFs, what ``proves'' that the
formalism 
is valid is our ability to perform global fits that include data from different
processes and at different energies. For the case of TMDs this has
been done only in a limited way. TMDs have been extracted from Drell-Yan and
$Z$-production data for values of $Q$ between 5 and 100 GeV (see, e.g.,
Refs.~\cite{Landry:2002ix,Konychev:2005iy,D'Alesio:2014vja}). 
In
semi-inclusive DIS, data from fixed-target experiments have played a truly
pioneering role for TMDs, but by themselves are not sufficient to check
the validity of the formalism.         
First of all, the range of $Q$ values is
limited, going from 1 to about 5 GeV. Secondly, there is a strong correlation
between the kinematic variables $Q$ and $x$. To study the effects of TMD
evolution, it is necessary to keep $x$ fixed and vary $Q^2$. So far, this has
been possible only for a few bins of COMPASS measurements. This has been
investigated in Ref.~\cite{Aidala:2014hva} and led to the conclusion that the
effects of TMD evolution are small in COMPASS data. To make the situation
worse, the data themselves are supposedly affected by some normalization error
that hamper our conclusion (see Erratum of Ref.~\cite{Adolph:2013stb}).

There is a first clear-cut check that needs to be done: verifying
the sign change of the Sivers function in Drell-Yan compared to semi-inclusive
DIS. This is one of the rare occasions when a sharp prediction has some
profound origin and leads to vast consequences. The prediction is based on
factorization theorems and the structure of Wilson lines in the definition of
TMDs. If this expectation were falsified, it would mean that there is
something very 
general that we do not understand about TMD factorization. In the less
dramatic scenario, it would mean that we misunderstand the nature
of the final-state interactions giving rise to T-odd effects, which we
presently attribute to soft light-cone gluon exchanges. In a more dramatic
scenario, the falsification of the sign-change prediction  
could demand
an extensive check also of what we normally take for granted not only in TMD
factorization, but also in collinear
factorization.   

If the Sivers function sign change is confirmed, this would give us confidence
in the validity of the TMD framework. It would however be just the beginning
of a long journey into the comprehension of single-spin asymmetries. 
These asymmetries are present not only in
the relatively clean context of semi-inclusive DIS and Drell-Yan twist-2
observables, but also in the historically famous $A_N$ asymmetries in
hadron-hadron collisions (see, e.g.,
Refs.~\cite{Abelev:2008af,Bland:2013pkt,Adamczyk:2013yvv,Drachenberg:2014txa,Drachenberg:2014jla}
to mention only the most recent ones, and the review by
Aschenauer, D'Alesio, Murgia in this Special Issue~\cite{Aschenauer:2015ndk}), in exclusive proton-proton collisions~\cite{Bultmann:2005na,Alekseev:2009zza,Adamczyk:2012kn}, in inclusive
pion electroproduction~\cite{Airapetian:2013bim,Allada:2013nsw}. It should be possible to explain these phenomena with
a single common language, but at the moment we just started scratching the
surface of this question. For hadron-hadron collisions, at high transverse
momentum we know that the appropriate language is that of collinear twist-3
factorization~\cite{Ji:2006vf,Kouvaris:2006zy,Kanazawa:2012kt,Kanazawa:2014dca,Kanazawa:2014nea}. 
We also know that there are connections between this
language and twist-2 TMD factorization. We are far from understanding to which
extent these connections are phenomenologically useful and survive QCD
corrections. In the meanwhile, $A_N$ asymmetries can be qualitatively
described in the
context of the so-called Generalized Parton Model~\cite{Anselmino:2012rq,Anselmino:2013rya,D'Alesio:2015uta}: there should be a
connection with the Wandzura-Wilczek approximation for
twist-3 DIS observables~\cite{Wandzura:1977qf}. However, in this approximation T-odd components are
normally discarded, in contrast with the Generalized Parton Model. Therefore,
if anything, the latter should be considered as a Wandzura-Wilczek
approximation with the addition of twist-2 T-odd terms.

An open problem from the point of view of the TMD formalism 
is twist-3 factorization. 
We know that there are unexpected 
 mismatches between
the results obtained with the TMD approach at twist 3 and the collinear
twist-3 approach~\cite{Gamberg:2006ru,Bacchetta:2008xw}. Maybe this is a signal of the impossibility to obtain
factorization at twist 3, but maybe this is an opportunity to push further
our knowledge of QCD and its technology.

The question of TMD factorization breaking in $pp$ collisions to hadrons demands
further scrutiny at all levels: formal, experimental, and
phenomenological. From the formal point of view, efforts should be made in
finding observables that are not affected by color entanglement problems, for
instance transverse-momentum-weighted or Bessel-weighted
quantities~\cite{Boer:2011xd}. On the
contrary, identifying observables that are clearly related to factorization
breaking would be extremely useful~\cite{Rogers:2013zha}. Note that if TMDs were accessible in
$pp$ collisions, a deeper treatment of the Wilson line would be required and
T-even functions could also become
nonuniversal~\cite{Buffing:2011mj,Buffing:2012sz,Buffing:2013kca,Buffing:2013dxa,Boer:2015kxa}. 
From the experimental point of view, the
challenge is to either find the signs of factorization breaking, or to
constrain their size. In order to do that, probably the only safe avenue is to
collect as much data as possible in $pp$ collisions, combine it with data from
other processes and look for tensions in phenomenological studies. 

From the experimental point of view, we need to extend the measurements in all
possible ways. For what concerns semi-inclusive data at fixed target
experiments, we are waiting for a new analysis of COMPASS multiplicities, for
the final publication of all measured azimuthal asymmetries from HERMES and
COMPASS, for further data from COMPASS, and for
the whole mass of results coming from JLab at 12 GeV. All of the measurements
should be done with multidimensional binning. Where possible, they should be
done for different targets (proton, deuterium, helium) and for
different final-state hadrons (pions, kaons, protons). 
 Attention should be paid to twist-3
observables. Beyond this, we look with great expectations to the construction
of an Electron Ion Collider (EIC), the ultimate machine for 3-D proton studies
(see the dedicated review by R. Ent in this Special Issue).

Even if we extend the range of kinematics with an EIC, semi-inclusive DIS will
not be sufficient by itself to disentangle the role of PDFs and FFs. 
Drell-Yan data will give an invaluable contribution. We look forward to having
results from COMPASS, Fermilab, Brookhaven, and even the LHC. COMPASS data
will give unprecedented information also about the 3-D structure of the pion,
the simplest of all hadrons, and yet only poorly known~\cite{Horn:2016rip}. Brookhaven, with his
unique feature 
in being the first and only polarized proton-proton collider, should provide
us with extended measurements of transverse single-spin asymmetries for the
production of jets, direct photons, and inclusive hadrons (sensitive to
twist-3 functions), for the production of $W$, $Z$, and Drell-Yan lepton
pairs (sensitive to TMDs), for the production of hadron pairs in the same jet
(sensitive to transversity), and finally for the production of two jets or two
hadrons in different jets (potentially affected by TMD factorization
breaking). Even if all the above facilities can perform experiments with
polarization, it should not be forgotten that we need
unpolarized data with the right characteristics for 3-D studies, in particular
transverse-momentum dependence and 
multidimensional binning.

To take full advantage of semi-inclusive DIS data, it is also essential to
obtain information about fragmentation functions and their 3-D
dependence (see the review by Garzia and Giordano in this Special Issue). 
High-luminosity electron-positron colliders are ideal to push these studies
forward. The single most important missing piece of information is at the
moment the transverse-momentum dependence of unpolarized fragmentation
functions. Priority should be given to measurements with sensitivity to
these quantities, with full multidimensional dependence ($z_1$, $z_2$,
$q_T$), if possible with different hadron types (all combinations of two
pions, pion-kaon, pion-proton, etc.). It is also important to scan different
values of center-of-mass energy. Future measurements at BELLE and
BESIII can perfectly suit these needs and will be complementary to each other
thanks to the different energies. A careful study of experimental acceptance
effects will be necessary, possibly calling for the need of Monte Carlo event
generators that include spin effects~\cite{Matevosyan:2011vj,Casey:2012ux}. 

From the phenomenological side, extractions of TMD PDFs and FFs are now
entering the so-called ``Phase 2,'' where the proper QCD definition of these objects is
taken into account and evolution equations are considered
The questions to answer are many: how do TMDs change as a function of $x$?
What are the differences between different flavors? What is the role of the
nonperturbative part of TMD evolution?

Last but not least, very little is known about gluon TMDs of all kinds, starting from the simplest
unpolarized one. Several interesting measurements have been proposed,
especially related to the gluon Boer-Mulders distribution, describing linearly
polarized gluons in an unpolarized
target~\cite{Boer:2010zf,Boer:2011kf,Boer:2012bt,Pisano:2013cya,Boer:2014lka,Dunnen:2014eta}. These
measurements could be 
carried out also at the LHC, or in the proposed fixed-target experiment at
LHC, AFTER~\cite{Anselmino:2015eoa}. 
Several challenging questions
need to be addressed, among which certainly the farthest reaching one is the
connection with the language of low-$x$ gluon TMDs (often called in this context unintegrated
distribution functions): the formalism at low $x$ is very different from that
of the standard TMDs, but it should be possible to find a common framework
that contains both versions of the formalism as limiting cases (see, e.g.,
Refs.~\cite{Marquet:2009ca,Dominguez:2011wm,Angeles-Martinez:2015sea,Boer:2016jnn}).

\section{GPDs}

In the last decade, a tremendous quantity of results related to GPDs was
obtained. Compared to TMDs, less activity from the formal side has taken
place. This is mainly due to the fact that the framework to analyze data and extract
GPDs was already laid out in a rigorous way (see the review by M.~Diehl in
this Special Issue~\cite{Diehl:2015uka}). 
However, several experimental
measurements have been published, by collaborations at DESY
and Jefferson Lab. The ``golden channel'' for GPD studies is
Deeply Virtual Compton Scattering (DVCS) (see the review by Kumericki, Liuti
and Moutarde in this Special Issue~\cite{Kumericki:2016ehc}). JLab Hall A
has measured the DVCS unpolarized and beam-polarized cross sections~\cite{Camacho:2006qlk},
CLAS measured beam spin asymmetries and longitudinally target spin asymmetries~\cite{Girod:2007aa,Pisano:2015iqa} 
and HERMES measured the complete set of beam charge, beam spin and target 
spin asymmetries~\cite{Airapetian:2001yk,Airapetian:2012mq,Airapetian:2010ab,Airapetian:2008aa,Airapetian:2011uq,Airapetian:2006zr,Airapetian:2009aa,Airapetian:2009cga}.
The DVCS unpolarized cross section has been measured also by the H1 and 
ZEUS collaborations~\cite{Chekanov:2003ya,Aktas:2005ty} at much higher energy
compared to fixed-target experiments. 

Complementary information
comes also from Deeply Virtual Meson Production (DVMP), where it is possible to probe
different combinations and different types of GPDs compared to DVCS (see the
review by Favart, Guidal, Horn and Kroll in this Special Issue~\cite{Favart:2015umi}). Moreover,
nucleon form factors are related to integrals of
GPDs. Therefore, form factor measurements indirectly constrain GPDs and have a
sharp impact on 3-D studies. Several parameterizations of GPDs have been presented
in the literature in the last years (see, e.g.,
Refs.~\cite{Guidal:2004nd,Goloskokov:2007nt,Goldstein:2013gra,Diehl:2013xca,Kumericki:2013br,Kumericki:2015lhb}
and see the detailed review by Guidal, Moutarde, and Vanderhaeghen in Ref.~\cite{Guidal:2013rya}). 

DVCS is and probably will remain the cleanest source of information about
GPDs. Even in this process, however, GPDs are not probed directly, but they
rather appear in Compton Form Factors, which are weighted integrals of
GPDs. In the past years, some efforts went into the extractions not of GPDs,
but rather of Compton Form Factors (see, e.g.,
Refs.~\cite{Guidal:2010de,Moutarde:2009fg,Kumericki:2013br,Boer:2014kya,Kumericki:2015tqa}). 
This approach has the advantage that it
does not require GPD modeling. However, it can only be considered as a step
toward the final goal of GPD extractions.

On top of this, 
GPDs depend on three variables, e.g., $H(x,\xi,t)$.\footnote{We remind the reader that
the so-called forward limit of GPDs corresponds to $H(x,0,0)$.}
It is practically impossible to scan their multidimensional
dependence. 
This problem, referred to as the ``curse of dimensionality''~\cite{Kumericki:2016ehc}, 
makes it extremely relevant to build models of GPDs
that are at the same time solid and flexible. They should fulfill all
fundamental properties of GPDs: this may seem an obvious statement, but its
implementation is not so obvious. Moreover, even if all constraints are
correctly incorporated, there will still be too much
freedom in the parameterization. Two extreme approaches can be followed: either
we accept to constrain only limited regions of GPDs (e.g., the cross-over
line, $x=\xi$),
or we resort to
models to extrapolate our knowledge beyond what is directly
accessible. Probably, the right approach sits in between these two extremes:
in addition to the formal requirements, 
some assumptions inspired by model calculations are required to limit the
flexibility of the parameterization and constrain the extrapolations. This
consideration holds true not only for GPDs, but in general for multi-dimensional
studies. The model assumptions should be well-motivated and the use of several
different models should help reaching more robust conclusions.

The interpretation of DVMP data is more challenging than DVCS data. This is
first of all due to the presence, beside GPDs, of meson Distribution
Amplitudes,
nonperturbative objects that are not well known, describing the probability amplitude to
find a $q \bar{q}$ pair. 
Secondly, there are some
difficulties in explaining the behavior of DVMP observables. The leading-twist
handbag diagram formalism predicts a well-defined behavior of the longitudinal
and transverse components of the cross section. Data, however, do not confirm these
predictions. The transverse cross section is typically larger than
expected in the region where data exist. The longitudinal cross section does decrease with $Q^2$ as
expected. These mismatches could be due to power and logarithmic corrections,
but this still needs to be
checked~\cite{Goloskokov:2006hr,Diehl:2007hd,Meskauskas:2011aa}. 
To understand the situation, it is
important to have a separation of longitudinal and transverse cross sections
for all the processes of interest. At the moment, such separation is not
available in most of pseudoscalar meson measurements. 

It was proposed to explain the large contribution of transverse photons in the
light meson channels in terms of transversity GPDs~\cite{Goloskokov:2009ia,Goloskokov:2011rd}. This explanation has extra
complications from the theoretical side, since it involves twist-3 pion wave
functions and the presence of transverse momentum~\cite{Martin:1999wb,Goloskokov:2007nt}, calling for a
transverse-momentum-dependent generalization of the established
formalism. Phenomenological estimates seem to be able to describe data. It
will be useful if these estimates can be put on stronger theoretical foundations.
 
In any case, the general conclusion is that DVMP observables, at least in experimentally accessible
kinematics, are more challenging to explain in terms of GPDs than DVCS. Future
data are expected from JLab after the 12 GeV upgrade. They will be useful to 
clarify some of the open issues and to understand to which extent DVMP can be
reliably used to extract GPDs. In the future, the EIC will reach higher $Q^2$
values and make it possible to explore also the contributions from sea quark and gluon GPDs. 

Table \ref{t:gpdtable} presents the full list of twist-2 GPDs, with their
dependence on polarization. Parametrizations in agreement with data 
are available for the first two
columns (chiral even
sector)~\cite{Guidal:2004nd,Goloskokov:2007nt,Goloskokov:2009ia,Goldstein:2010gu,Goldstein:2013gra,Diehl:2013xca,Kumericki:2013br,Kumericki:2015lhb}. These parameterization make use of subsets of all
available GPD data. Some of them rely only on the knowledge of the collinear
PDFs and the nucleon Form Factors, and obtain the full GPD on the basis of
some specific choice of functional form.

\renewcommand{\arraystretch}{1.5}
\begin{table}
\begin{center}
\begin{tabular}{c|c|c|c|c|c|}
\multicolumn{2}{c}{}&\multicolumn{4}{c}{quark pol.}\\
\cline{3-6}
\multicolumn{2}{c|}{}& U & L & \multicolumn{2}{c|}{T} \\ 
\cline{2-6}
\multirow{3}{*}{\rotatebox{90}{nucleon pol.}} &{U} &
{ $\boldsymbol{H}$}   & & \multicolumn{2}{c|}{$E_T+2 \tilde{H}_T$} \\
\cline{2-6}
& {L} & &{ $\boldsymbol{\tilde{H}}$} & \multicolumn{2}{c|}{ $\tilde{E}_T$} \\
\cline{2-6}
& {T} & { $E$} &  { $\tilde{E}$} &
{$\boldsymbol{H_T}$} & {${\tilde{H}_T}$} \\
\cline{2-6} 
\end{tabular}
\end{center}
\caption{Twist-2 Generalized Parton Distribution functions (GPDs).
The U,L,T correspond to unpolarized, longitudinally polarized and transversely
polarized nucleons (rows) and quarks (columns). Functions in boldface survive
in the forward limit.}
\label{t:gpdtable}
\end{table}
 
Different GPD extractions/models are based on different assumptions and functional
forms. The first class of fits is based on the so-called Double Distributions
(DDs), introduced by Radyushkin~\cite{Radyushkin:1998es} and M\"uller~\cite{Mueller:1998fv}.
DDs are convenient ways to represent GPDs and automatically fulfill some fundamental properties of GPDs, such as
polynomiality (related to Lorentz invariance). 
Examples of parameterizations based on DDs are the
Vanderhaeghen-Guichon-Guidal VGG model~\cite{Guidal:2004nd}
and the Goloskokov-Kroll (GK)
model~\cite{Goloskokov:2007nt,Goloskokov:2009ia}. 
The DD models in general reproduce the data fairly well. They however
underestimate the unpolarized DVCS cross section, they do not describe very
well the beam spin
asymmetry data at low $t$ and they have problems reproducing some of the
azimuthal moments~\cite{Guidal:2013rya}. The GK model is also able to describe some DVMP observables.

Another family of GPD parameterization is based on the use of conformal
moments. In the
so called ``dual
parameterization,'' GPDs are expanded on a series of Gegenbauer
polynomials, which makes it easier to apply evolution
equations. Phenomenological studies based on the dual parameterization have
been published in Refs.~\cite{Guzey:2005ec,Guzey:2006xi,Guzey:2008ys,Polyakov:2008xm}. Qualitatively, the results are the same as for the
DD parameterizations.
Another conformal-moment parameterization uses the Mellin--Barnes
representation of GPDs, and has been worked out by Kumericki and M\"uller in
Refs.~\cite{Kumericki:2009uq,Kumericki:2011zc,Kumericki:2013br}. The
parameterization in Ref.~\cite{Kumericki:2013br} is at present the one
that can describe most of available DVCS data in a satisfactory way. One of the
main differences between this extraction and the others is the presence of a
large $\tilde{H}$ GPD.

A fourth type of GPD parameterization is based on model results. The
parameterization of Ref.~\cite{Goldstein:2010gu} is based on a spectator model
for the nucleon, with additional Regge-inspired flexibility. The agreement with
data is similar to that of the DD parameterization.

The plots in the recent experimental paper of Ref.~\cite{Pisano:2015iqa} give
a good overall idea of the ability of the various parameterizations to describe
new data: the situation is qualitatively good, especially if we consider that the
observables to describe are diverse and most
of the parameterization are based on simple concepts with few free parameters.

Apart from the details of fitting techniques, GPDs can be used to reconstruct
parton density maps in a two-dimensional transverse position space and a one
dimensional longitudinal momentum space.  
Already from the study of Form Factors and
their Fourier transform, we can obtain transverse maps in impact parameter space, integrated
over longitudinal momentum, and only for valence quark combinations~\cite{Guidal:2004nd,Diehl:2013xca}. From the
Fourier transform of the Dirac form factors, it turns out that the
distribution of valence
up quarks is narrower than the down. The root-mean-square impact parameter is about 0.7 fm. 
Among other things, this means that a high-energy probe sees a core of
positive charge in the center of the proton and a cloud of negative charge
around it. 
In a transversely polarized proton, we know that the densities of up and down quarks are
distorted in opposite ways, and the distortion of down quarks is larger than
for up quarks. This distortion indirectly suggests that the up quarks have a
large orbital angular momentum opposite to the proton spin. Vice-versa for the
down quarks. 

When considering also the longitudinal momentum dependence, it turns out that
at high $x$ the impact-parameter distribution of down quarks seems to be wider
than up (root-mean-square impact parameters are equal to 
0.8 fm and 0.4 fm respectively)~\cite{Diehl:2013xca}. At low $x$, the width of the distribution becomes
wider and can even diverge, in the same way for up and down quarks. These
features were obtained from the study of form factors together with PDFs,
but assumptions are needed to connect the two limits of the parent GPD.
A first attempt to obtain this information directly from Compton Form Factor
measurements was illustrated in Ref.~\cite{Guidal:2013rya}. Assuming that the
Compton Form Factor can give us direct information about $H(x,x,t)$, the
reconstruction of impact-parameter distributions requires an extrapolation to
the forward limit $H(x,0,t)$ (also called ``deskewing'' correction), which is
model dependent, and a Fourier-Bessel transform about $t$. The outcome of this
study confirms that the the width of the impact-parameter distribution widens
at lower $x$.

Thanks to collider measurements at HERA, it is possible also to extract the
GPDs for sea quarks and gluons. The outcome of the fits is that there is no
drastic difference between the behavior of sea quarks and gluons. Gluons are
slightly narrower than sea quarks. Their
root-mean-square impact parameters are 0.7-0.9 fm and 0.6-0.8 fm, respectively. Another observation is that the width
in impact parameter space for sea quarks and gluons has no strong dependence
on the longitudinal variable $x$~\cite{Kumericki:2009uq,Goldstein:2010gu}.

\section{Wigner distributions}

An alluring frontier to reach is the possibility of experimentally accessing
Wigner distributions~\cite{Ji:2003ak,Belitsky:2003nz,Lorce:2011kd} or at least their Fourier transform, generalized TMDs
(GTMDs) \cite{Meissner:2009ww,Meissner:2008ay,Lorce:2013pza}. There are 16 complex-valued
GTMDs at leading twist. In the forward limit, they reduce to the eight TMDs of
Tab.~\ref{t:TMDtable} and upon integration over transverse momentum they
reduce to the eight GPDs of Tab.~\ref{t:gpdtable}. 
It would already be of extreme interest
to identify a process that
can be 
used {\em in principle}, even if it
turns out to be too difficult to realize in practice. A theoretical QCD
analysis of this process should lead to factorization theorems involving Wigner
distributions and to their evolution equations, which should eventually
match onto the TMD and GPD evolution equations in their respective
limits. Very recent work in this direction was presented in Ref.~\cite{Hatta:2016dxp,Echevarria:2016mrc}.

Even if we are eventually able to extract/reconstruct Wigner distributions,
there is still an important clarification to make. Referring to TMDs or GPDs, 
we speak routinely about
3-D imaging or 3-D mapping. We should remember that we are talking
about ``probability densities,'' not about complete images. 
Reconstructing a
digital image means knowing exactly the color of each pixel, given the color
of all others, not just knowing the probability that a pixel in a certain
position is red. Information analogous to real images corresponds to
multiparton correlation functions, which tell us what is the probability of
finding a parton in a certain condition given the conditions of the other
partons (ideally, all of them). Multiparton correlation functions have been the subject of intense
theoretical and experimental studies in the last years (see, e.g.,
~\cite{Gaunt:2009re,Calucci:2010wg,Diehl:2011yj,Strikman:2010bg,Rinaldi:2013vpa}).

\section{Nucleon spin decomposition}
 
A fundamental problem to address is the composition of nucleon's
spin. Without entering too much into the details of the study of parton
angular momentum, let us summarize here some general considerations (for more
details, see the
dedicated review by Lorc\'e and Liu in this Special Issue~\cite{Liu:2015xha}). There are
various equally valid definitions of partonic total and orbital angular
momentum, generally falling into the two classes of kinetic and canonical
angular momentum,
connected to the two historical approaches of Ji~\cite{Ji:1996ek} and Jaffe-Manohar~\cite{Jaffe:1989jz},
respectively. 
It is in principle possible and interesting to ``measure'' Orbital Angular
Momentum (OAM) in both
definitions. The
difference between the two can tell us something nontrivial about QCD
dynamics. In practice, it would already be an historic achievement to measure
OAM in one of the two definitions. 

A full decomposition of the nucleon's angular momentum means that we should
measure spin and
total angular momentum, or spin and orbital angular momentum, for all quark
flavors and gluons. Even assuming a flavor-blind sea ($\bar{u}=\bar{d}=s=\bar{s}$), this means we should
measure eight contributions. It will be extremely hard if not impossible to
check the validity of the spin sum rule (i.e., that the sum of all
contributions adds up to $1/2$). We will probably assume the validity of the
sum rule and use it to determine those contributions that are harder to
access. Therefore, even 
measuring only some of the contributions remains a valuable
achievement. 

We should also keep in mind that the various contributions to total
spin can occur with different signs. Therefore, if we determine that the
sum of certain contributions is small, it does not mean that all the
contributions in the sum are small. They can be large and opposite.
There has been considerable excitement in the last year concerning the fact
that the contribution of the gluon spin is 
maybe sufficient to saturate the spin sum
rule~\cite{deFlorian:2014yva,Nocera:2014gqa}. This estimate is affected by large uncertainties, and still leaves much
room for a large net contribution of orbital angular momentum. In any case,
even if it will remain true that spin contributions add up to something close
to $1/2$,
we still have to check and justify why orbital angular momenta from different
partons add up to zero. 

Another important observation is that we would like to know also the angular
momentum {\em density}, i.e., as a function of $x$ (and $k_T$ and $b_T$). The
situation in this case becomes more complex and also frame-dependent. What we
have learned for sure up to now is that the integrand of Ji's relation ($x
H(x,0,0)+ x E(x,0,0)$) cannot be interpreted as an angular momentum
density~\cite{Leader:2012ar,Hatta:2012jm,Harindranath:2013goa,Leader:2013jra}. 
Orbital angular momentum can be defined starting from Wigner distributions,
since they contain information about both $k_T$ and
$b_T$~\cite{Lorce:2011ni,Hatta:2011ku}. Depending on the choice of Wilson line
in the definition of the Wigner distribution, it was shown that both
definitions of
orbital angular momentum can be recovered: the kinetic definition using a
straight Wilson line, the canonical definition using a Wilson line of the same
shape as in the definition of TMDs~\cite{Ji:2012sj}. It
was pointed out that the difference between the two results can be connected
to a ``chromomagnetic torque'' experienced by the active quark due to the
effect of final state interactions~\cite{Burkardt:2012sd}.
Furthermore, the kinetic definition can be related to 
twist-3 collinear
GPDs~\cite{Penttinen:2000dg,Kiptily:2002nx,Hagler:2003jw,Hatta:2012cs,Lorce:2015lna,Courtoy:2013oaa}. The
connection between the Wigner distribution definition and the collinear
definition has been clarified in Ref.~\cite{Rajan:2016tlg}, which also
suggests a connection to the pure twist-3 part of the structure function $g_2$, accessible in
experimental measurements.

For the canonical definition, at present only the definition in terms of a
 Wigner distribution is available~\cite{Lorce:2011ni,Hatta:2011ku}. However,
 as for all other
Wigner distributions, no way to
access it experimentally has been proposed.

At the moment, the only practical way to extract partonic total angular
momentum is to use Ji's relation~\cite{Ji:1996ek}. As is well known, the relation contains two
terms, one of which is well known, being the second Mellin moment of the
unpolarized collinear PDFs, i.e., the partonic fractional linear momentum. The
second piece is obtained by integrating the forward limit of the GPD
$E(x,\xi,t)$. Note, however, that some model calculations showed some inconsistencies, still not
completely understood, when total angular momentum is split into various
flavors according to Ji's relation~\cite{Lorce:2011kd,Liu:2014zla,Ji:2015sio}.

In any case, there is evidence from
GPD and TMD studies that total angular
momentum (in the kinetic definition) for up quarks is large and positive,
while for down quarks is small and
negative~\cite{Guidal:2004nd,Ahmad:2006gn,Goloskokov:2008ib,Bacchetta:2011gx,Diehl:2013xca}. Rough\-ly
speaking, this leaves room for about 50\% of the proton spin to be carried by gluons. 
The latest
lattice QCD calculations are in partial agreement with these indications: they
also observe a large and positive total angular momentum for up quarks and a
small and negative total angular momentum for down quarks. However, there is
the unexpected indication that orbital angular momentum for sea quarks is
large~\cite{Deka:2013zha}, compensated by the fact that gluons carry only
less than 30\% of the proton spin. 
It should be kept in mind that these lattice
calculations are done on a quenched lattice and require extrapolations to the
physical pion mass. Unquenched calculations typically increase the momentum
fraction of gluons and therefore they might also increase the fraction of
angular momentum. Extrapolation to the physical pion mass may cause an
overestimate of the up quark angular momentum.

\section{Further topics}

Even if this topical review is devoted to the 3-D structure of the proton,
there is still plenty of interesting physics to measure and understand in the
1-D case, or even in the 0-D case. For instance, 
we know the electric charge of the proton very well (even though we still
do not know why it has that value), but other ``charges'' can be defined: the
axial, tensor, and scalar charges. 

Charges represent one-dimensional information about the proton. They could be
interesting in two ways, as is always the case with hadronic quantities:
they could reveal something new about QCD, and they can be used to reveal
something new about other fields of research, through processes that involve
hadrons. Low-energy precision measurements of such processes can provide
unique probes of new physics at much higher scales.
For example, 
there is the possibility
that nucleon beta decay or the neutron electric dipole moment receive contributions from interactions
not included in the Standard Model. The knowledge of the nucleon charges is necessary to extract
the corresponding hypothethical couplings from precision
measurements.
At present, this information is taken from lattice QCD, but in the future it should be
possible to obain it from experimental measurements (see, e.g., Refs.~\cite{Cirigliano:2013xha,Bhattacharya:2015esa,Courtoy:2015haa}).
 
The axial charge is related to the integral
of helicity PDF, the tensor charge to integrals of the
transversity PDF, and the scalar charge to the integral of the twist-3 PDF
$e(x)$~\cite{Schweitzer:2003uy,Ohnishi:2003mf,Cebulla:2007ej,Mukherjee:2009uy,Lorce:2014hxa}. The
latter can be accessed, for instance, in 
semi-inclusive DIS~\cite{Efremov:2002ut,Bacchetta:2003vn,Mao:2013waa,Courtoy:2014ixa,Gohn:2014zbz}. 

In several points of our discussion higher-twist parton distributions were
mentioned. This topic by itself can become a focus theme in the next
years. It is unavoidably present in all measurements, but it is notoriously
difficult to disentangle and interpret. The first step is therefore to
separate higher-twist contributions from the rest (see, e.g.,
Refs.\cite{Jimenez-Delgado:2013boa,Monfared:2014nta,}). 
To make it possible, it is
desirable to have the largest possible $Q^2$ span, keeping all
other kinematics as fixed as possible. At the same time, studies of peculiar
twist-3 signals (e.g., specific modulations in the cross section that vanish
at twist 2) can provide
unique information on the physics of the nucleon and challenge our
theoretical interpretation. The $e(x)$ twist-3 PDF mentioned
above represents a good example, but much more can be studied.

As we discussed at the beginning, understanding the proton means that we know
how to compute its characteristics, reproducing what is measured and
predicting what it is not. The most successful approach so far has been that
of lattice QCD. 
In the past, lattice QCD was able to study only specific hadronic
quantities. In the field of 3-D structure, lattice QCD traditionally provides only calculations of
Mellin moments of PDFs and GPDs. This is already an important achievement, and
it is already nontrivial to obtain results in agreement with experimental
information, as we have already discussed for the case of the tensor charge.

A step forward into lattice studies of 3-D nucleon structure was performed by
pioneering investigations on
TMDs \cite{Musch:2010ka,Hagler:2009mb,Musch:2011er,Engelhardt:2015xja}. 
This approach makes it possible to calculate transverse moments of TMDs and
leads also to a lattice-based understanding of the effect of different
Wilson-line choices, including the difference between kinetic and canonical
partonic angular momenta.

Recently, studies have appeared that suggest
a way to compute on the lattice the full functional dependence of parton
distributions, including 3-D
ones~\cite{Ji:2013dva,Xiong:2013bka,Ji:2014hxa,Ji:2015jwa,Ji:2015qla}. 
The calculation is performed in a frame
with finite longitudinal momentum (producing the so-called ``quasi PDFs'') 
and the connection to light-front parton
distributions is performed using perturbative matching conditions. This
approach is potentially extremely powerful and opens up the possibility of
calculating the structure of the nucleon instead of extracting it from
data. There is however still a very long way ahead: present estimates of
1-D PDFs are very crude and not superior to any basic model
calculation~\cite{Chen:2016utp}. Problems of different nature affect both the high-$x$ and low-$x$
region. Nevertheless, improvements in computational efficiency and in the
understanding of quasi PDFs could in the future lead us to a stage comparable
to {\em a priori} calculations of electronic orbitals.



\begin{acknowledgement}
Discussions with F.~Delcarro, L.~Mantovani, B.~Pasquini, C.~Pi\-sa\-no, M.~Radici,
X.~Xiong are gratefully acknowledged. 
This work is supported by the European Research Council (ERC) under the European Union's 
Horizon 2020 research and innovation programme (grant agreement No. 647981, 3DSPIN).
\end{acknowledgement}

\bibliographystyle{epjB}
\bibliography{mybiblio}

\end{document}